# Local Electric Field Measurement in GaN Diodes by exciton Franz-Keldysh Photocurrent Spectroscopy

Darpan Verma,[1] Md Mohsinur Rahman Adnan,[2] Mohammad Wahidur Rahman,[2] Siddharth Rajan,[2] and Roberto C. Myers[1,2]*

[1] *Department of Materials Science Engineering, The Ohio State University, Columbus, Ohio, 43210 USA*
[2] *Department of Electrical and Computer Engineering, The Ohio State University, Columbus, Ohio, 43210, USA*
**myers.1079@osu.edu*

**The eXciton Franz-Keldysh (XFK) effect is observed in GaN p-n junction diodes via the spectral variation of photocurrent responsivity data that redshift and broaden with increasing reverse bias. Photocurrent spectra are quantitatively fit over a broad photon energy range to an XFK model using only a single fit parameter that determines the lineshape, the local bias ($V_l$), uniquely determining the local electric field maximum and depletion widths. As expected, the spectrally determined values of $V_l$ vary linearly with the applied bias ($V$) and reveal a large reduction in the local electric field due to electrostatic non-uniformity. The built-in bias ($V_{bi}$) is estimated by extrapolating $V_l$ at $V = 0$, which compared with independent C-V measurements indicates an overall ±0.31 V accuracy of $V_l$. This demonstrates sub-bandgap photocurrent spectroscopy as a local probe of electric field in wide bandgap diodes that can be used to map out regions of device breakdown (hot spots) for improving electrostatic design of high voltage devices.**

GaN is a III-V compound semiconductor with a wide bandgap ($E_g$~3.4 eV) and a high breakdown electric field ( > 3.3 MV/cm), making it useful for power devices and optoelectronics.[1–5] Heterostructures based on GaN result in high responsivity Metal-Heterojunction-Metal (MHM) ultraviolet photodetectors.[6,7] However, the stability and lifetime at high voltages for GaN based power electronics still remains a challenge to the device industry.[7–12] Vertical devices such as p-n junctions have electric field peaks near the device edge, which are managed using junction termination structures such as field plates and guard rings.[13–16] Lateral devices such as AlGaN/GaN HEMTs also show field variation across the gate-drain region. The built-in electric field in such heterostructures can be measured using contactless electroreflectance,[17,18] however it is not suitable at high fields when the reflectance oscillations decay. It is a challenge to measure the actual field distribution in such devices, and device engineers usually rely on device simulations to estimate the electric field profiles. Furthermore models used to estimate electric field profiles may ignore inhomogeneity, and thus may not provide accurate estimates of hot spots in such structures.[19–22] Thus, a local probe of electric field would be a useful tool for mapping out the field distribution, identifying hot spots, and validating or refining complex electrostatic models and device designs.

Semiconductors exhibit the Franz-Keldysh (F-K) effect due to which photons with energy ($E_{ph}$) below the bandgap ($E_g$) are absorbed due to electric-field induced band bending. Electron and hole wave functions exhibit finite overlap at energies below the bandgap due to band bending, which results in a sub-bandgap absorption tail.[23] The band bending can be characterized by photocurrent spectroscopy through the redshift of the absorption spectrum with reverse bias.[24–26] Recently, the electric field maximum, depletion width, and surface charge in AlGaN/GaN heterostructures were determined based on fitting photocurrent and photovoltage spectra to a tunneling based WKB model derived for the case of linear and parabolic band bending.[27] Unfortunately, this model is not applicable to the field profile in p-n diodes. In another approach, the F-K effect in a GaN p-n junction and Schottky diodes were modeled to determine the electric field in the device active region.[28–30] In analyzing their results, Maeda *et al.* used the Franz-Keldysh Aspnes (FKA) model[31] to calculate the $V$ dependence of photocurrent measured at constant excitation wavelength taking into account the field-dependence of the band-to-band absorption.

Building on these previous works, here we develop a spectrally-resolved measurement and quantitative model of the F-K effect in GaN p-n diodes. The photocurrent responsivity data show a spectral variation that cannot be described by only the band-to-band FKA effect.[31] Instead the lineshape indicates an eXciton Franz-Keldysh (XFK) effect [32–36]. We derive a quantitative fitting procedure that utilizes normalized photocurrent responsivity data that are insensitive to position-dependent variation in light intensity and optoelectronic efficiency. Quantitative fits of the normalized photocurrent responsivity spectra to the XFK model are obtained using only a single adjustable parameter that determines the lineshape, $V_l$, the local voltage drop across the pn-diode. This spectrally determined parameter uniquely determines the local electric field maximum and depletion widths, thus demonstrating a pathway to achieve sensitive electric field mapping in wide bandgap devices.

The GaN p-n junction is grown by plasma-assisted molecular beam epitaxy (PAMBE) in a Veeco Gen 930 system. A schematic of the GaN p-n diode is shown in Fig 1. The layers are grown under Ga-rich conditions at a substrate temperature (T$_{sub}$) of 675°C (pyrometer), at a growth rate of 4 nm/min, RF power of 300 W, and N$_2$ flow rate of 2.5 sccm. The epitaxial structure for the diode consists of 1 μm n- GaN drift layer (Si = 1×10$^{17}$ cm$^{-3}$) grown on a 4-μm thick n+ GaN/sapphire template. Next a 200 nm thick layer of p- GaN (Mg = 1×10$^{18}$ cm$^{-3}$) is grown at T$_{sub}$ = 650°C. Finally, the 30 nm p+ GaN cap layer (Mg = 1×10$^{20}$ cm$^{-3}$) is grown at T$_{sub}$ = 600°C. The Si and Mg concentration are estimated based on secondary ion mass spectroscopy (SIMS) doping calibration. The top ohmic contact to p+ GaN is formed by depositing a Pd/Ni/Au (30 nm/30 nm/30 nm) metal stack using an e-beam evaporator followed by annealing in N$_2$ for 1 min at 400°C. The p-GaN layer is etched for mesa isolation in ICP RIE using BCl$_3$/Cl$_2$. Following a deeper etch, the n$^+$ GaN layer is contacted using an Al/Ni/Au (30 nm/30 nm/150 nm) metal stack deposited by e-beam evaporation. The average net doping concentration $(N_d - N_a)$ determined by C-V measurement was $8\,(\pm 0.4) \times 10^{16}$ cm$^{-3}$ consistent with background C acceptor compensation of 2×10$^{16}$ cm$^{-3}$ in the MBE. Figure 1(a) shows the band edge diagram, obtained using a one-dimensional self-consistent Poisson solver.[37,38] The carrier concentration in the two regions is shown in Fig. 1(b), where the zero-bias depletion region width is 0.12 μm.

The photocurrent spectroscopy setup, as shown in Figure 1(d) incorporates reflective optics (Al-coated) enabling measurement across the deep-UV to visible wavelengths without chromatic aberration. The light source is a 75W Xenon Arc lamp (OBB Powerarc) focused through the entrance slit of a 140 mm monochromator (Horiba MicroHR) with a 250 nm blazed holographic grating. The average spectral bandwidth across the measurements reported here is 2.5 nm. An optical chopper operating at 200 Hz (2.5 ms open/close time) modulates the monochromatic light, which is focused to a position (x,y) on the device using a 40× reflective microscope objective with a theoretical spot diameter ~0.5 μm over the excitation wavelength range. The photocurrent is pre-amplified and then fed into a digital lock-in amplifier (Zurich instruments HFLI), referenced to the chopper frequency. Reverse bias ($V < 0$) is applied to the device using a Keithley 2604B source meter unit. The photocurrent is normalized by the average optical power (measured after the microscope objective) at each excitation wavelength using a wavelength corrected power meter (Thorlabs PM 100D) to obtain $I_{PR}$, the photocurrent responsivity spectrum as a function of photon energy ($E_{ph}$) at various applied biases ($V$), Fig. 2(a). The data show a characteristic exciton absorption peak in GaN that redshifts with applied bias, as well as the redshift and broadening of the sub-band gap absorption tail. Next, we derive a quantitative model for the field-dependent photocurrent spectra in GaN based on the XFK effect.

The excitation is tightly focused to position (x, y) using a reflective microscope objective at the top of the device with an incident photon flux $\varphi_0$. The reflectance of GaN over the measurement range (~3-3.5eV) is relatively constant (~0.11-0.13),[39] however at each position the reflectance of the surface varies especially over the top metal electrode. Thus, the photon flux entering the p-GaN layer is reduced to $\varphi_0(1-r)$. Within the p-GaN the photon flux is further reduced to $\varphi_0(1-r)e^{-\alpha_p L_p}$ before photons reach the depletion region, where $\alpha_p$ is the doping broadened absorption spectrum of p-GaN and $L_p$ is the thickness of the flatband p-GaN region. As the minority electron diffusion length is >200 nm,[40] all photocarriers produced in the 200-nm thick p-GaN region contribute to photocurrent at any bias. Similarly, given the electric fields present in the depletion region and the electron and hole mobilities of GaN, all photocarriers in the depletion region are assumed to contribute to photocurrent. After passing through the depletion region, the photon flux entering the flatband n-GaN region is $\varphi_0(1-r)e^{-[\alpha_p L_p + \int_{-W_p}^{W_n} \alpha(F(z))dz]}$, where $W_n$ ($W_p$) is the depletion width of the n(p) region and $\alpha(F(z))$ is the field-dependent absorption coefficient of GaN. The photon flux remaining after passing through the photocarrier collection region, is $\varphi_0(1-r)e^{-[\alpha_p L_p + \alpha_n L_n + \int_{-W_p}^{W_n} \alpha(F(z))dz]}$, where the minority hole collection is restricted to within a distance $L_n$ of the edge of the depletion region and $\alpha_n$ is the doping broadened absorption spectrum of n-GaN. Because the GaN/sapphire back interface has r~0.14-0.2 and the photon flux is also reduced by a factor of ~0.33 due to beam divergence over the thickness of the n-GaN template layer, the back reflected photon flux is <1/10 the remaining flux. Thus, back reflected light can be disregarded. The photocurrent density is then simply the photon absorption inside the collection region, $J = e(1-r)\varphi_0 \left[1 - a_{pn} e^{-\int_{-W_p}^{W_n} \alpha(F(z))dz}\right]$, where $a_{pn} = e^{-[\alpha_p L_p + \alpha_n L_n]}$ is the absorption occurring in the flatband portion of the photocarrier collection region. Dividing the photocurrent density by the incident power ($E_{ph} \varphi_0$), we obtain an expression for the $F(z)$ and $E_{ph}$ dependent photocurrent responsivity in units of A/W,

$$I_{PR} = \frac{e(1-r)}{E_{ph}} \left[1 - a_{pn} e^{-\int_{-W_p}^{W_n} \alpha(F(z))dz}\right]. \quad (1)$$

This derivation accounts for surface reflection, minority carrier collection outside of the depletion regions, and the optical path of the experiment as described above. Accordingly, the $I_{PR}$ spectral lineshape is sensitive to electric field via $\alpha(F(z))$. In our measurements, the p-n diode is kept in reverse-bias-mode. The dark current is removed using lock-in detection, and thus, the measured current is due only to photons absorbed in the collection region that are converted into electron-hole pairs, collected by the electrodes. Because the diode is not in forward bias mode, stray electric fields cannot energize carriers that might induce different transport mechanisms that would otherwise complicate the analysis.[41] As a result, the $I_{PR}$ spectral lineshape inherently depends on local electrostatics and provides a way to map the local electric field across devices. The Aspnes model was used to describe bias dependent photocurrent (I-V) measurements in GaN Schottky and pn-diodes measured at constant $E_{ph}$.[28,29] However, in modeling the spectral ($E_{ph}$) dependence of $I_{PR}$, we found the FKA prediction to qualitatively disagree with lineshape over a wide $E_{ph}$ range (see Supplementary Material); the $I_{PR}$ spectral dependence is indicative of the exciton FK effect that is qualitatively distinct from the non-excitonic FKA theory.

Exciton absorption in GaN is observed at room temperature with an excitonic absorption peak of $E_{ph}$ = 3.44 eV at zero field, where GaN has an F = 0 exciton binding energy of $E_X^0$ = 20.4 meV.[42] As pointed out by Dow and Redfield,[32] Blossey,[43] and Merkulov,[33] the Coulombic electron-hole interaction modifies the absorption spectrum (exciton absorption), which has a qualitatively distinct field-dependence compared with the band-to-band transition treated by the FKA theory. This exciton-modified FK effect was previously observed in GaN electroabsorption spectra.[34–36] As our data span a wide range of photon energy deficit, $\Delta = \frac{E_g - E_{ph}}{E_X}$, including the range where $E_g - E_{ph}$ is comparable to the exciton binding energy ($E_X$), the values of the absorption coefficient are greatly underestimated by the non-excitonic FKA model; when $\Delta \sim 1$ the electric field distortion of the exciton wave function enhances the absorption coefficient (see Supplementary Material for comparison of FKA versus XFK spectral lineshape). To model the XFK effect, we utilize the theory developed by Merkulov,[33] where the sub-band gap field-dependent absorption coefficient is,

$$\alpha(F(z)) = \frac{Cx}{\pi^2(\delta^2 x^2 + 1)} \;;$$

$$\delta = \Delta - 1 - \frac{9f^2}{2} \;;\quad x = \frac{8}{f} e^{\left(-\frac{4\Delta^{3/2}}{3\;f} - \frac{2}{\sqrt{\Delta}} \ln\left(\frac{8\Delta^{3/2}}{f}\right)\right)} \;;\quad f = \frac{eF(z)a}{E_X}, \quad (2)$$

where C is the exciton wave function normalization coefficient, $f$ is the electric field in dimensionless units, $a$ is the Bohr radius, the quadratic Stark effect shifted exciton binding energy is $E_X = E_X^0 + bF(z)^2$, and $b = \frac{9e^2a^2}{8E_X^0}$ is the exciton polarizability.[44]

The vertical (z-axis) electric field profile at position (x,y) is,[41]

$$F(x,y,z) = \begin{cases} \frac{-eN_a}{\epsilon_s}(W_p(x,y) + z), & -W_p(x,y) \leq z < 0 \\ \frac{-eN_d}{\epsilon_s}(W_n(x,y) - z), & 0 < z \leq W_n(x,y) \end{cases}, \quad (3)$$

where, $\epsilon_s = 10.4\epsilon_0$ is the static dielectric constant of GaN, $N_a$ ($N_d$) is the acceptor (donor) density, $W_n(x,y)$ ($W_p(x,y)$) is the depletion width of the n (p) region at position (x,y) given by,

$$W_n(x,y) = \sqrt{\frac{-2\epsilon_s V_l(x,y)N_a}{q(N_d+N_a)N_d}} \quad , \quad W_p(x,y) = \sqrt{\frac{-2\epsilon_s V_l(x,y)N_d}{q(N_d+N_a)N_a}}, \quad (4)$$

where $V_l(x,y)$ is the total local vertical bias at position (x,y). Note that the total bias is typically assumed to be the sum of the applied and built-in bias, $V_{total} = V - V_{bi}$, with the built-in bias defined as a positive quantity. However, as discussed below, the total local bias can differ greatly due to electrostatic non-uniformity (field fringing) mainly related to the electrode geometry. Thus, we formatted the equations above to emphasize the local nature of $V_l(x,y)$ measurements obtained by spectral measurements.

Within a p-n junction, Eq. (3) describes the triangular electric field profile in the depletion region. The field slope is independent of bias as it is a function of doping density, which is determined from C-V measurements (see Supplementary Material). The local field maximum is affected by changes in the local depletion widths ($W_n$ and $W_p$), which, in turn, depend on the local bias ($V_l(x,y)$). As follows from Eqs. (3-4), $W_n(x,y)$, $W_p(x,y)$, and $F(x,y,z)$ are uniquely determined by $V_l(x,y)$. It is therefore the only independent variable determining the field-dependent absorption term in Eq. (1), $\int_{-W_p}^{W_n} \alpha(F(z))dz$.

To facilitate spectral lineshape fits, the responsivity data are normalized by their value at an energy well above the band gap ($E_{ph}^0 = 3.6\ eV$) taken at $V = 0$, at which $V_l = -V_{bi} = -(k_BT/e)\ln(N_aN_d/n_i^2) \sim -3.2\ V$, where $k_B$ is Boltzmann's constant, T is the temperature, and $n_i$ as the intrinsic carrier concentration, $10^{-10}$ cm$^{-3}$. This yields the unitless responsivity, $I_{PR}^N$, which following from Eq. (1) is given by,

$$I_{PR}^N = \frac{E_{ph}^0}{E_{ph}}\left[1 - a_{pn}e^{-\int_{-W_p}^{W_n} \alpha(F(z))dz}\right] \quad (5).$$

The field-independent absorption spectrum, $a_{pn}$, is determined from the $I_{PR}^N$ data at V = 0 for which $\int_{-W_p}^{W_n} \alpha(F(z))dz \sim 0.00$, thus,

$$a_{pn} = 1 - \frac{E_{ph}}{E_{ph}^0} \cdot I_{PR}^N(V=0) \quad (6).$$

The normalization in Eqs. (5-6) eliminates the dependence of $I_{PR}$ on r and bias-independent absorption (see derivation in Supplementary Material). The measured $I_{PR}^N$ spectra are plotted in Fig. 2(b) on a semi-logarithmic scale at various values of V. These data are fit to Eq. (5) noting that all parameters are experimentally measured or known except for $\int_{-W_p}^{W_n} \alpha(F(z))dz$. As noted earlier, $V_l$ is the only free parameter determining the depletion widths and field profile in the latter, and therefore determines the spectral dependence of $I_{PR}^N$ through the XFK effect, $\alpha(F(z))$ of Eq. (2). From a search through the literature we were unable to find a calculated value for the C parameter of Eq. (2) for GaN, however, as it is a simple $E_{ph}$ independent scaling term, its value can be estimated from published absorption spectra of GaN from the literature. Fitting data from Refs. [42,45] we find an average C = 9·10$^7$ (see Supplementary Materials). This estimate is further refined by performing two parameter fits of the measured $I_{PR}^N$ spectrum at V = -40 V, varying both C and $V_l$, from which we find C = 6.3·10$^7$ (see Supplementary Material).

Holding C as a constant, the $I_{PR}^N$ spectra are fit using only $V_l$ as a free parameter. We obtain excellent agreement between the data and model (R$^2 \geq 0.998$). Error bars (precision) of the $V_l(x, y)$ values are calculated based on the standard error of the spectral fits, which varies from ±0.125 to ±2 V. The values of $V_l(x, y)$ obtained from spectral fits are plotted as a function of V in Fig. 3(a), revealing a linear relation. We therefore obtain the empirical relation characterizing the total local bias,

$$V_l(x, y) = c_l(x, y)V - V_{bi}, \quad (7)$$

where $c_l(x, y)$ is a unitless coefficient of proportionality. The local effective applied bias ($V_l(x, y) + V_{bi}$) is reduced by a factor of $c_l(x, y)$ compared with the average applied bias. Thus, $c_l(x, y)$ characterizes the local electric field inhomogeneity. The extrapolated y-axis intercept ($V = 0$) of $V_l(x,y) = -V_{bi} = -2.89 \pm 0.23\ V$ in Fig. 3(a), which matches the independently determined value of $V_{bi} = 3.08 \pm 0.02\ V$ obtained via C-V measurements, Fig. 3(b). Theoretically, $V_{bi} = -3.2$ V, which matches the $V_l(x,y)$ estimate within 0.31 V, but differs from the C-V measured value by 0.19 V. Thus, the overall accuracy of the set of $V_l(x,y)$ data is estimated to be $\pm 0.31$ V.

Finally, in Fig. 4 the average z-axis electric field profile for V = -40 V is plotted and compared with the local electric field profile determined from fitting the $I_{PR}^N$ spectrum. At location (x,y) of the $I_{PR}$ measurements, the field maximum and depletion width are far smaller than the average, indicating that this location is an electrostatic cold spot. The bias dependent spectral measurement and fits described above are reproduced at a second location to verify the reproducibility of the electric field measurement method. At a different location for a larger device, we measure another electrostatic cold spot (see Supplementary Material). Thus, this method of determining local $F(z)$ by fitting the spectral variation of $I_{PR}$ can prove useful in designing GaN based p-n devices within breakdown constraints by mapping out the local electric field and identifying cold and hot spots. Our results highlight the importance of accurately determining local field variation particularly in high-voltage devices where the difference between the local and average vertical field magnitudes can be greatly magnified.

**Supplementary Material**

See online supplementary material that includes derivation of the normalized photocurrent responsivity fit equation, comparison of the XFK and FKA effect models, estimate of exciton wavefunction normalization parameter, and additional photocurrent responsivity spectra taken at a different location on the device, together with spectral fitting and local electric field measurement.

**Acknowledgments** Funding for this research was provided by the Center for Emergent Materials: an NSF MRSEC under award number DMR-1420451 and by the AFOSR GAME MURI (Grant FA9550-18-1-0479, Program Manager Dr. Ali Sayir).

# References

[1] M. Meneghini, G. Meneghesso, and E. Znoni, *Power GaN Devices: Materials, Applications and Reliability* (Springer International Publishing Switzerland, 2016).

[2] J.C. Zolper, Solid. State. Electron. **42**, 2153 (1998).

[3] Y. Yoshizumi, S. Hashimoto, T. Tanabe, and M. Kiyama, J. Cryst. Growth **298**, 875 (2007).

[4] S. Rajan, P. Waltereit, C. Poblenz, S.J. Heikman, D.S. Green, J.S. Speck, and U.K. Mishra, IEEE Electron Device Lett. **25**, 247 (2004).

[5] H., Amano, Y., Baines, E, Beam, M. Borga, T., Bouchet, P.R., Chalker, M., Charles, K.J., Chen, N., Chowdhury, R., Chu, C., De Santi, M.M. De Souza, S., Decoutere, L., Di Cioccio, B., Eckardt, T., Egawa, P., Fay, J.J., Freedsman, L., Guido, G., Haynes, T., Heckel, D., Hemakumara, P., Houston, J., Hu, M., Hua, Q., Huang, A., Huang, S., Jiang, H., Kawai, D., Kinzer, M., Kuball, A., Kumar, K.B., Lee, X., Li, D., Marcon, M., Marz, R., McCarthy, G., Meneghesso, M., Meneghini, E., Morvan, A., Nakajima, E.M.S., Narayanan, S., Oliver, T., Palacios, P., Daniel, M., Plissonnier, R., Reddy, M., Sun, I., Thayne, A., Torres, N., Trivellin, V., Unni, M.J., Uren, M.V., Hove, D.J., Wallis, J., Wang, J., Xie, S., Yagi, S., Yang, C., Youtsey, R., Yu, E., Zanoni, S., Zeltner, Y., and Zhang, J. Phys. D. Appl. Phys. **51**, 163001 (2018).

[6] H.Z. Xu, Z.G. Wang, M. Kawabe, I. Harrison, B.J. Ansell, and C.T. Foxon, J. Cryst. Growth **218**, 1 (2000).

[7] Y. Tian, S.J. Chua, and H. Wang, Solid. State. Electron. **47**, 1863 (2003).

[8] J. Wuerfl, E. Bahat-Treidel, F. Brunner, E. Cho, O. Hilt, P. Ivo, A. Knauer, P. Kurpas, R. Lossy, M. Schulz, S. Singwald, M. Weyers, and R. Zhytnytska, Microelectron. Reliab. **51**, 1710 (2011).

[9] P.J. Martinez, E. Maset, E. Sanchis-Kilders, V. Esteve, J. Jordán, J.B. Ejea, and A. Ferreres, Semicond. Sci. Technol. **33**, 45006 (2018).

[10] J.A. Del Alamo and E.S. Lee, IEEE Trans. Electron Devices **66**, 4578 (2019).

[11] G. Meneghesso, G. Verzellesi, F. Danesin, F. Rampazzo, F. Zanon, A. Tazzoli, M. Meneghini, and E. Zanoni, IEEE Trans. Device Mater. Reliab. **8**, 332 (2008).

[12] H. Ishida, D. Shibata, H. Matsuo, M. Yanagihara, Y. Uemoto, T. Ueda, T. Tanaka, and D. Ueda, in *2008 IEEE Int. Electron Devices Meet.* (2008).

[13] H. Fu, K. Fu, S.R. Alugubelli, C. Cheng, X. Huang, H. Chen, H. Yang, C. Yang, J. Zhou, S.M. Goodnick, and F.A. Ponce, IEEE Electron Device Lett. **41**, 1 (2020).

[14] C.B. Goud and K.N. Bhat, IEEE Trans. Electron Devices **38**, 1497 (1991).

[15] H. Ohta, K. Hayashi, F. Horikiri, T. Nakamura, and T. Mishima, Jpn. J. Appl. Phys. **57**, 1 (2018).

[16] J.R. Laroche, F. Ren, K.W. Baik, S.J. Pearton, B.S. Shelton, and B. Peres, J. Electron. Mater. **34**, 370 (2005).

[17] C. Wetzel, H. Amano, and I. Akasaki, J. Appl. Phys. **85**, 3786 (1999).

[18] Y.T. Hou, K.L. Teo, M.F. Li, K. Uchida, H. Tokunaga, N. Akutsu, and K. Matsumoto, Appl. Phys. Lett. **76**, 1033 (2000).

[19] D. Nagulapally, R.P. Joshi, and A. Pradhan, AIP Adv. **5**, 17103 (2015).

[20] S. Verma, S.A. Loan, and A.M. Alamoud, J. Comput. Electron. **17**, 256 (2018).

[21] J. Möreke, C. Hodges, L.L.E. Mears, M.J. Uren, R.M. Richardson, and M. Kuball, Microelectron. Reliab. **54**, 921 (2014).

[22] J. Luo, S.L. Zhao, M.H. Mi, W.W. Chen, B. Hou, J.C. Zhang, X.H. Ma, and Y. Hao, Chinese Phys. B **25**, 2 (2015).

[23] A. Cavallini, L. Polenta, M. Rossi, T. Stoica, R. Calarco, R.J. Meijers, T. Richter, and H. Lüth, Nano Lett. **7**, 2166 (2007).

[24] J. Ruo-Lian, W. Jun-Zhuan, C. Peng, A. -, A. MultilayerStructure on Si Jiang Ruo-Lian, Z. Zuo-Ming, and C. Pen, Chinese Phys. B **12**, 785 (2003).

[25] S. Wang, T. Li, J.M. Reifsnider, B. Yang, C. Collins, A.L. Holmes, and J.C. Campbell, IEEE J. Quantum Electron. **36**, 1262 (2000).

[26] S.C. Shen, Y. Zhang, D. Yoo, J.B. Limb, J.H. Ryou, P.D. Yoder, and R.D. Dupuis, IEEE Photonics Technol. Lett. **19**, 1744 (2007).

[27] Y. Turkulets and I. Shalish, J. Appl. Phys. **124**, 75102 (2018).

[28] T. Maeda, T. Narita, M. Kanechika, T. Uesugi, T. Kachi, T. Kimoto, M. Horita, and J. Suda, Appl. Phys. Lett. **112**, 252104 (2018).

[29] T. Maeda, M. Okada, M. Ueno, Y. Yamamoto, M. Horita, and J. Suda, Appl. Phys. Express **9**, 91002

(2016).

[30] T. Maeda, T. Narita, H. Ueda, M. Kanechika, T. Uesugi, T. Kachi, T. Kimoto, M. Horita, and J. Suda, Appl. Phys. Lett. **115**, (2019).

[31] D.E. Aspnes, Phys. Rev. **147**, 554 (1966).

[32] J.D. Dow and D. Redfield, Phys. Rev. B **1**, 3358 (1970).

[33] I. Merkulov, Sov. J. Exp. Theor. Phys. **66**, 2314 (1974).

[34] F. Binet, J. Duboz, E. Rosencher, F. Scholz, and V. Härle, Phys. Rev. B - Condens. Matter Mater. Phys. **54**, 8116 (1996).

[35] J.Y. Duboz, F. Binet, E. Rosencher, F. Scholz, and V. Härle, Mater. Sci. Eng. B **43**, 269 (1997).

[36] M.A. Jacobson, O. V Konstantinov, D.K. Nelson, S.O. Romanovskii, and Z. Hatzopoulos, *J. Crys. Gro. 230, 459-461*. (2001).

[37] M. Grundmann, BandEng: Poisson-Schrodinger Solver Software (2004).

[38] B. Jogai, J. Appl. Phys. **91**, 3721 (2002).

[39] G. Yu, G. Wang, H. Ishikawa, M. Umeno, T. Soga, T. Egawa, J. Watanabe, and T. Jimbo, Appl. Phys. Lett. **70**, 3209 (1997).

[40] K. Kumakura, T. Makimoto, N. Kobayashi, T. Hashizume, T. Fukui, and H. Hasegawa, Appl. Phys. Lett. **86**, 1 (2005).

[41] S.M. Sze and K.K. Ng, *Physics of Semiconductor Devices Third Edition* (Wiley-Interscience, 2007).

[42] J.F. Muth, J.H. Lee, I.K. Shmagin, R.M. Kolbas, H.C. Casey, B.P. Keller, U.K. Mishra, and S.P. DenBaars, Appl. Phys. Lett. **71**, 2572 (1997).

[43] D.F. Blossey, Phys. Rev. B **2**, 3976 (1970).

[44] G. Weiser, Phys. Rev. B **45**, 14076 (1992).

[45] O. Ambacher, W. Rieger, P. Ansmann, H. Angerer, T.D. Moustakas, and M. Stutzmann, Solid State Commun. **97**, 365 (1996).

# Figures Captions

**Fig. 1.** Schematic of PAMBE grown GaN p-n junction diode with its corresponding (a) Band edge diagram along the z-axis, where $E_C$, $E_V$ and $E_f$ are the conduction band edge, valence band edge and fermi energy, respectively. (b) Acceptor ($N_a$) and donor ($N_d$) doping density profile, and (c) z-component of the xy-averaged electric field ($F_{avg}$) calculated using Eqs. 3-4 substituting $V_l(x, y, z)$ with $(V - V_{bi})$. (d) The photocurrent spectroscopy setup and device geometry.

**Fig. 2.** Photocurrent responsivity ($I_{PR}$) as a function of incident photon energy ($E_{ph}$) for various values of applied bias ($V$) plotted on a linear scale. (b) Normalized photocurrent responsivity ($I_{PR}^N$) spectra (points) measured at several applied biases ($V$) compared with fits (lines) to Eq. 5, where the field-dependent absorption coefficient is given by the XFA theory of Eq. 2. The local voltage ($V_l$) is adjusted to achieve the best fit to the data and quantify the local field profile at the location (x,y).

**Fig. 3.** (a) $V_l$ as a function of $V$ (points) obtained from fitting the $I_{PR}^N$ spectra to the XFK model. The color of each point matches that of the spectra of $I_{PR}(E_{ph})$ from Fig. 2 for the different values of $V$. Error bars are determined from the standard error of the spectral fits. The line is a fit to Eq. 7, such that the intercept indicates $-V_{bi}$. (b) Shows (1/C$^2$) in units of ($10^{17}$ F$^2$/m$^4$) plotted versus V where the linear intercept indicates $V_{bi}$.

**Fig.4.** The local z-component of the electric field ($F$) along the z-axis at location P(x,y) of the photocurrent spectral responsivity measurements at $V = -40\ V$ as determined from quantitative fits of the photocurrent responsivity spectra to the XFK model. The local profile is compared with the expected average electric-field profile of the GaN p-n junction diode at an applied bias of -40V.

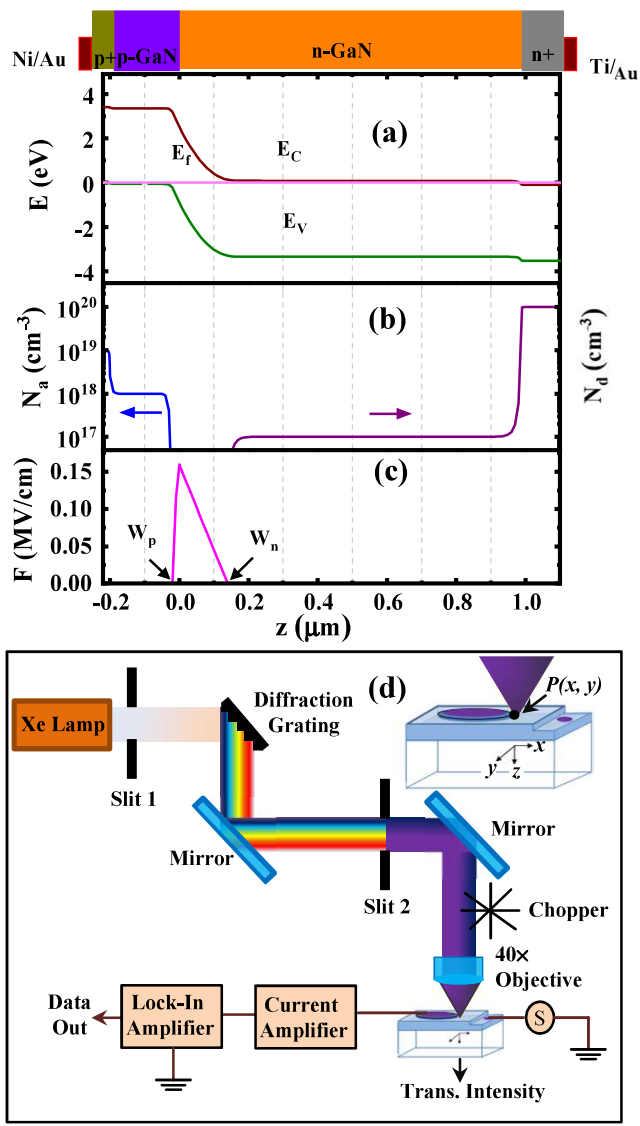

Fig. 1

Verma et al.

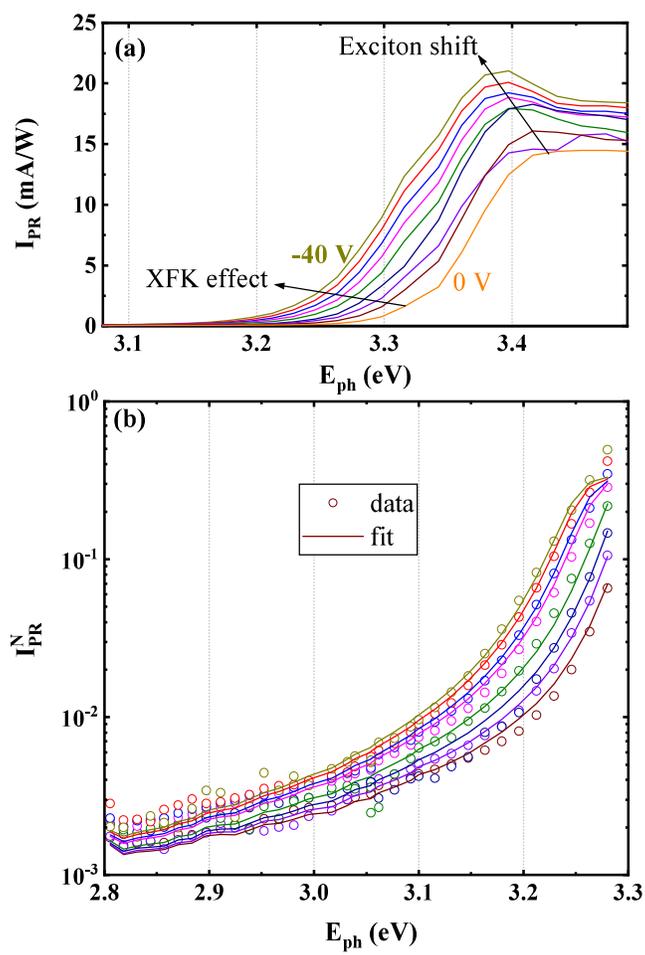

Fig. 2

Verma et al.

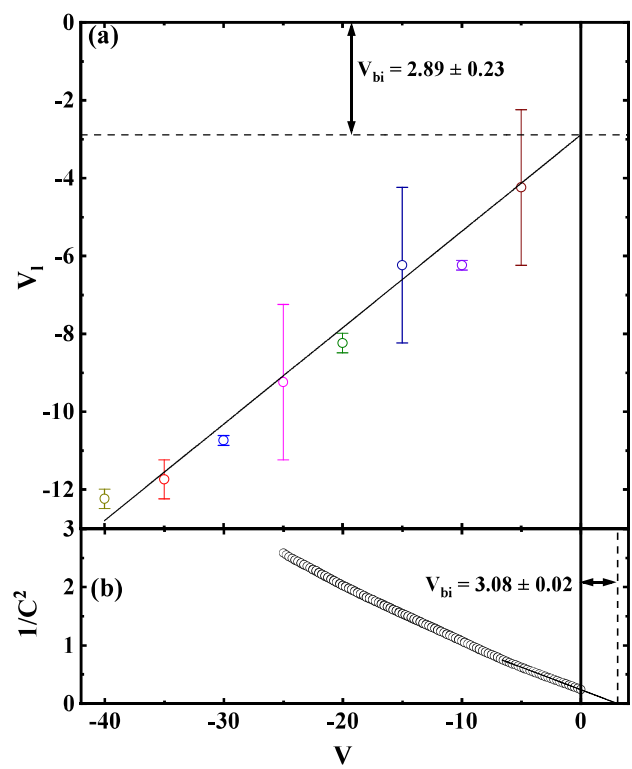

Fig. 3

Verma et al.

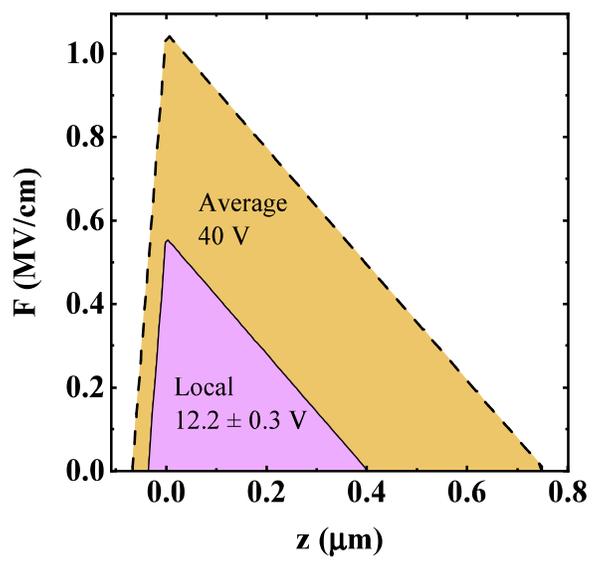

Fig. 4

Verma et al.

Supplementary materials

# Local Electric Field Measurement in GaN Diodes by exciton-modified Franz-Keldysh Photocurrent Spectroscopy

Darpan Verma,[1] Md Mohsinur Rahman Adnan,[2] Mohammad Wahidur Rahman,[2] Siddharth Rajan,[2] and Roberto C. Myers[1,2*]

[1)] Dept. of Materials Science Engineering, The Ohio State University, Columbus, Ohio, 43210 USA
[2)] Dept. of Electrical and Computer Engineering, The Ohio State University, Columbus, Ohio, 43210, USA
*myers.1079@osu.edu

**Normalized photocurrent responsivity and spectral fit function derivation**

The photocurrent responsivity derived in units of (A/W) as a function of $E_{ph}$ and $V_l$, (equation (1) in the main text), is given by,

$$I_{PR}(E_{ph}, V_l) = \frac{e[1-r]}{E_{ph}}\left[1 - a_{pn}(E_{ph})e^{-\int_{-W_p(V_l)}^{W_n(V_l)} \alpha(E_{ph}, F(z,V_l))dz}\right] \quad (S1),$$

where $a_{pn}(E_{ph}) = e^{-[\alpha_p(E_{ph})L_p + \alpha_n(E_{ph})L_n]}$ and compared with Eq. (1) of the main text, here we explicitly state the $V_l$ dependence of the depletion widths ($W_n(V_l)$ and $W_n(V_l)$), and z-dependent field profile ($F(z, V_l)$). This derivation considers photocarrier collection in the flat-band-portions of the n and p regions due to minority carrier diffusion as well as the field dependent absorption in the depleted region. As described in the main text, the field sensitivity of the spectra comes from the term, $\int_{-W_p(V_l)}^{W_n(V_l)} \alpha\left(E_{ph}, F(z, V_l)\right) dz$, which depends on the position dependent parameter $V_l(x, y)$, the total local bias. However, the terms r and $a_{pn}(E_{ph})$ can also be position dependent, thus data cannot be fit to Eq. (S1) without utilizing several independent fit parameters. In order to reduce the analysis to a single fit parameter of the spectral lineshape, a normalization procedure is derived that eliminates the position dependent parameter r, and provides a means to determine the parameter $a_{pn}(E_{ph})$. The analysis requires that at least two photocurrent spectra be acquired at each position, with V = 0 and V < 0.

Following from Eq. (S1), the photocurrent spectrum at V = 0 ($V_l = -V_{bi}$) is given by,

$$I_{PR}(E_{ph}, -V_{bi}) = \frac{e[1-r]}{E_{ph}}\left[1 - a_{pn}(E_{ph})e^{-\int_{-W_p(-V_{bi})}^{W_n(-V_{bi})} \alpha(E_{ph}, F(z, -V_{bi}))dz}\right].$$ Under these weak field conditions, the absorption in the depletion region is negligible, $\int_{-W_p(-V_{bi})}^{W_n(-V_{bi})} \alpha(E_{ph}, F(z, -V_{bi}))dz$ $\cong 0.00$ for $E_{ph} < E_g$. We find that within the Merkulov model[1], this assumption is numerically valid in GaN for $E_{ph} < 3.30\ eV$, yielding,

$$I_{PR}(E_{ph}, -V_{bi}) = \frac{e(1-r)}{E_{ph}}[1 - a_{pn}(E_{ph})] \quad (S2).$$

This V = 0 photocurrent spectrum is normalized by its value at $E_{ph}^0 > E_g$, where $I_{PR}(E_{ph}^0, -V_{bi}) = \frac{e(1-r)}{E_{ph}^0}\left[1 - e^{-[\alpha_p(E_{ph}^0)L_p + \alpha_n(E_{ph}^0)L_n]}\right]$. If $E_{ph}^0 > E_g$, then $\alpha_p(E_{ph}^0) \sim \alpha_n(E_{ph}^0) > 10^5$ cm$^{-1}$. Because $L_p + L_n > 400$ nm, then $e^{-[\alpha_p(E_{ph}^0)L_p + \alpha_n(E_{ph}^0)L_n]} = e^{-[>4]} < 0.02$. Therefore,

$$I_{PR}(E_{ph}^0, -V_{bi}) \cong \frac{e(1-r)}{E_{ph}^0} \quad (S3).$$

Normalizing the photocurrent by this value,

$$a_{pn}(E_{ph}) \cong 1 - \left(\frac{E_{ph}}{E_{ph}^0}\right) I_{PR}^N(E_{ph}, -V_{bi}) \quad (S4),$$

which is Eq. (6) of the main text. Note that the right-hand side of this equation is fully determined from the measured photocurrent spectrum at V = 0, $I_{PR}(E_{ph}, -V_{bi})$.

As described in the main text, the photocurrent spectra at V < 0 ($V_l < -V_{bi}$) are normalized by $I_{PR}(E_{ph}^0, -V_{bi})$ to define the unitless responsivity spectral function, which following from Eqs. (S1-S4) can be written as,

$$I_{PR}^N(E_{ph}, V_l) = \frac{E_{ph}^0}{E_{ph}}\left[1 - \left(1 - \left(\frac{E_{ph}}{E_{ph}^0}\right) I_{PR}^N(E_{ph}, -V_{bi})\right) e^{-\int_{-W_p(V_l)}^{W_n(V_l)} \alpha(E_{ph}, F(z, V_l))dz}\right] \quad (S5),$$

where we explicitly state the $E_{ph}$ and $V_l$ dependences compared with Eq. 5 of the main text.

Note this normalization eliminates the positional dependence of r from the analysis. As the term, $a_{pn}(E_{ph})$ is fully determined experimentally using Eq. (S4), then the only free parameter in Eq. (S5) is $V_l$, and all other variables are experimentally measured. Thus, S5 can be used to model the $I_{PR}^N(E_{ph}, V_l)$ spectral data using Eq. (2), where the depletion width and field profile is determined from Eqs. (3,4), i.e. $W_{n|p}(V_l) \propto \sqrt{V_l}$, $F(z, V_l) \propto \sqrt{V_l} \pm z$.

# Exciton wave function normalization coefficient, C

Within Merkulov's XFK model of the field dependent absorption coefficient,

$$\alpha(F(z)) = \frac{Cx}{\pi^2(\delta^2 x^2 + 1)} \;;$$

$$\delta = \Delta - 1 - \frac{9f^2}{2} \;; \quad x = \frac{8}{f} e^{\left(-\frac{4\Delta^{3/2}}{3\,f} - \frac{2}{\sqrt{\Delta}} \ln\left(\frac{8\Delta^{3/2}}{f}\right)\right)} \;; \quad f = \frac{eF(z)a}{E_X} \quad (S6)$$

Within this theory, the non-excitonic FK-effect can be approximated, as derived by Merkulov,

$$\alpha(F(z))_{KF} = \frac{C}{32\pi^2} \frac{\Delta}{f} e^{\left(-\frac{4\Delta^{3/2}}{3\,f}\right)}, \quad (S7)$$

in which $\Delta = \frac{E_g - E_{ph}}{E_X}$, $E_X = E_X^0 + bF(z)^2$ is the exciton binding energy taking into account the quadratic Stark shift in which the polarizability is given by $b = \frac{9e^2 a^2}{8 E_X^0}$, and a is the Bohr radius. Equation (S7) is applicable in the range that is far away from the excitonic peak at 3.44eV. The coefficient, C, is the same in both Eq. S6 and S7. All GaN specific parameters at room temperature ($E_g$=3.39 eV, $E_X^0$=20.4 meV) are known from literature except C. Here we describe fits of the Merkulov model of $\alpha(F(z))$ to absorption spectra in literature from Refs. [2-3]. Noting that Eqs. (S6) and (S7) give the absorption coefficient, which varies exponentially with $E_{ph}$ near $E_g$, data are fit on a semi-logarithmic scale, such that,

$$\log(\alpha(F(z)) = \log(C) + \log\left(\frac{x}{\pi^2(\delta^2 x^2 + 1)}\right), \quad (S8)$$

$$\log(\alpha(F(z)_{KF}) = \log(C) + \log\left(\frac{1}{32\pi^2} \frac{\Delta}{f} e^{\left(-\frac{4\Delta^{\frac{3}{2}}}{3\,f}\right)}\right), \quad (S9)$$

where log(C) only impacts the data as an amplitude correction, a vertical shift of $\log(\alpha(F(z)))$ or $\log(\alpha(F(z)_{KF})$.

Figure S1 plots fits of Eq. (S6) and (S7) to absorption data from the literature.[2,3] The C parameter is listed in each. The values of $\alpha$ in the literature are quite varied in the sub-band gap region owing to variation in the crystalline quality, doping, and Urbach tail. Nevertheless, in data taken close to the exciton peak in absorption,[3] the Merkulov KF model of absorption agrees when

$C = 1\times10^8$. In data taken below the exciton peak,[2] the Merkulov KF model of absorption agrees when $C = 8\times10^7$ Taking the average of these values of C, we find $C = 9\times10^7$.

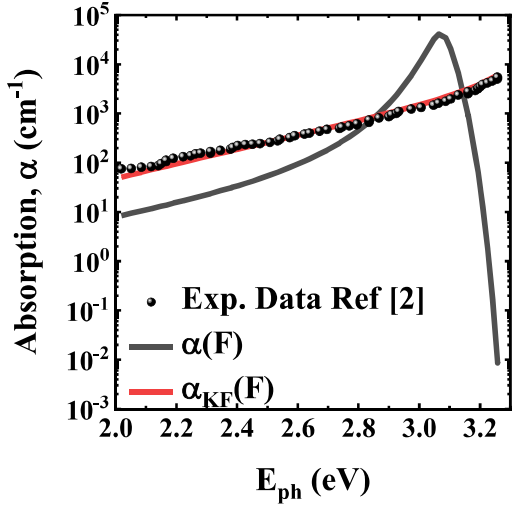 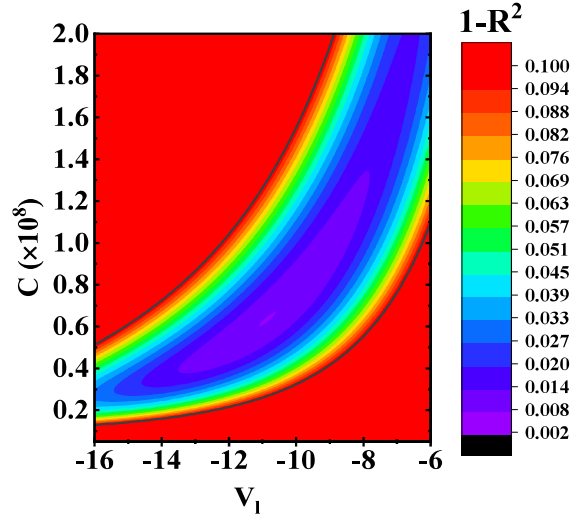

Figure S1. *Absorption spectrum of GaN fit to Merkulov KF model.*

Figure S2. *Variation of $R^2$ from spectral fits of $I_{PR}^N$ at V = -40 V as a function of C and $V_l$.*

The value of C is further refined by carrying out two parameter fits of the $I_{PR}^N$ data at V = -40 V, to Eq. (S5) where sub-band gap XFK tail is most pronounced. Figure S2 plots the (1-$R^2$) value of $I_{PR}^N$ determined at different values of C and $V_l$. There is a unique minimum at $C = 6.3 \times 10^7$ (with $R^2$>0.998) as expected due to the independence of these parameters under the measured conditions indicating the validity of the two-parameter fit. Figure S3(a) shows the resulting best fit (line) of the V = -40 V data (points), with the resultant fit parameters values of $C = 6.3 \times 10^7$ and $V_l = 12.23\ V$. In Fig. S3(b) and (c), the sensitivity of obtained parameters can be seen by varying the value of $V_l$ (best fit $\pm\ 2\ V$) and $C$ (best fit $\pm\ 2 \times 10^7$), respectively. Clearly, variation in C adjusts the amplitude at all $E_{ph}$, whereas $V_l$ impacts mainly the lineshape (redshift and broadening) changes associated with the XFK-effect.

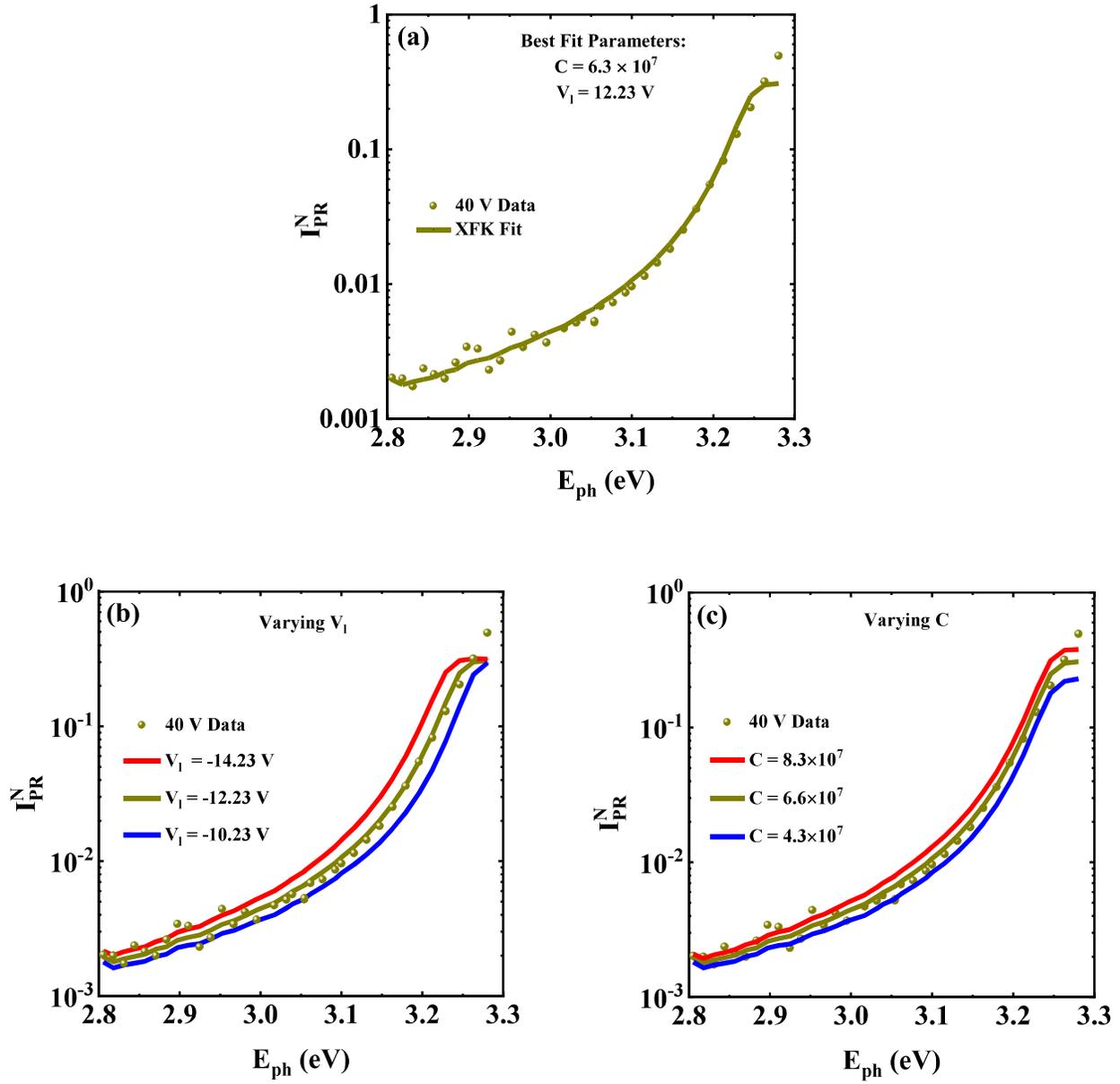

*Figure S3.* $I_{PR}^{N}$ *spectral data from V = -40 V fit to the XFK model using two-parameter fits. (a) The best fit of C and $V_l$ after sweeping through paremeter space. (b) Effect of variation in $V_l$ on the spectral fit quality, and (c) effect of C variation on fit quality.*

**Comparison of Exciton-free and Exciton-modified FK photocurrent spectra**

As mentioned in the main text, we found the exciton-free Franz-Keldysh-Aspnes (FKA) [4] model unable to match our data over a broad photon energy range. Aspnes derived an analytic expression for the electric field dependence of the absorption coefficient for direct gap semiconductors, the FKA-model,[4]

$$\alpha_{FKA}(E_{ph}, F(z)) = \frac{|M_{cv}|^2 \mu_\perp}{\epsilon_0 m_0^2 n c E_{ph}} \left(\frac{2\mu_\| e^7 F(z)}{\hbar^5}\right)^{\frac{1}{3}} \left[Ai'^2\left(\frac{E_g - E_{ph}}{\left(\frac{e^2\hbar^2 F(z)^2}{2\mu_\|}\right)^{\frac{1}{3}}}\right) - \frac{E_g - E_{ph}}{\left(\frac{e^2\hbar^2 F(z)^2}{2\mu_\|}\right)^{\frac{1}{3}}} Ai^2\left(\frac{E_g - E_{ph}}{\left(\frac{e^2\hbar^2 F(z)^2}{2\mu_\|}\right)^{\frac{1}{3}}}\right)\right],$$

(S8)

where $\mu = \frac{m_e m_h}{m_e + m_h}$ is the reduced mass, $\mu_\| = 0.12 m_0$ ($\mu_\perp = 0.16 m_0$) is the reduced mass along (perpendicular to) the z-component of the electric field, $m_e$ ($m_h$) is the electron (hole) effective mass, $m_0$ is the electron rest mass, $n = 2.5$ is the refractive index of GaN assumed constant over the data fitting range of $E_{ph} = 2.8 - 3.3\ eV$, $c$ is the speed of light, $\hbar$ is the reduced Planck's constant, $e$ is the electron charge, $\epsilon_0$ is the vacuum permittivity, and $|M_{cv}|^2 \sim E_g m_0^2 / 2\mu_\perp \sim 2 \times 10^{-48}\ J \cdot kg$ is the momentum matrix element. This model was used to describe the bias dependent photocurrent (I-V) measurements in GaN Schottky and pn-diodes measured at constant $E_{ph}$.[5,6] However, we found that the FKA model gives a qualitatively different spectral lineshape of $I_{PR}(E_{ph})$ compared with the data. We illustrate these results by fitting $I_{PR}^N(E_{ph}, V_l)$ to equation (S1) using $V_l$ as the only free parameter, where $\alpha(E_{ph}, F(z))$ is given by either the Aspnes model $\alpha_{FKA}(E_{ph}, F(z))$ as given by Eq. (S8), or the Merkulov exciton-modified FK model $\alpha(E_{ph}, F(z))$ as given by Eq. (S6).

Figure S4(a) plots $I_{PR}^N$ as a function of $E_{ph}$ taken at $V = -40\ V$. Data are plotted as points, and fits are plotted as lines. Clearly the FKA model shows a linear spectral variation on a log scale, whereas the data show a distinct slope change at higher photon energies reflecting an increase in absorption compared with what the FKA model predicts. This behavior was previously observed in GaN and discussed in Refs.[7–9]. The Merkulov model can account for this excitonic FK effect quite clearly. Figure S4(b-h) show the same results for fits taken at lower biases. At all biases, we find the XFK model valid and the FKA model in strong disagreement. Figure S5 shows the $V$ vs $V_l$ estimate for Aspnes model which gives an unphysical $V_{bi} = 15.2 \pm 2.5\ V$.

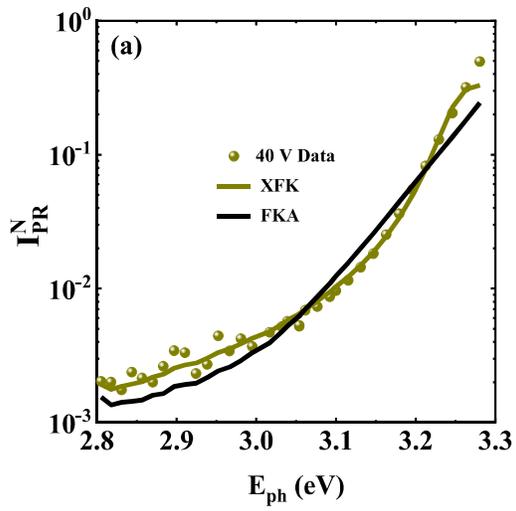
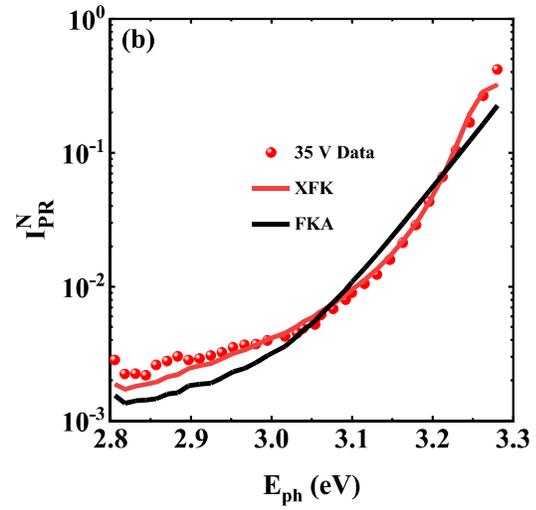
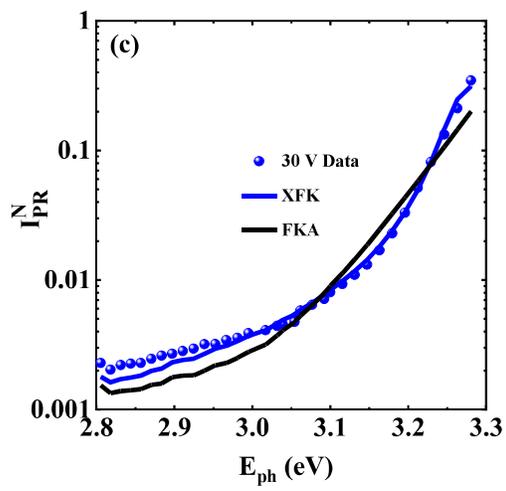
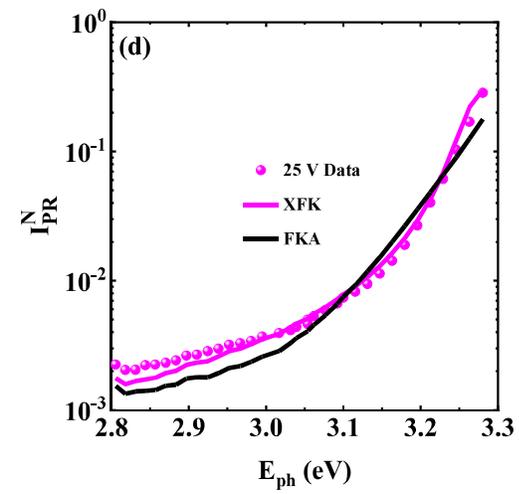
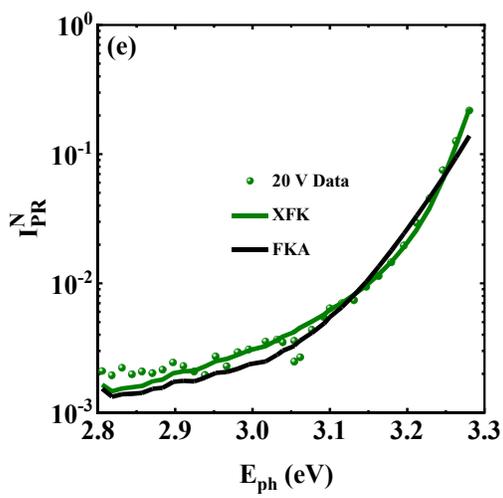
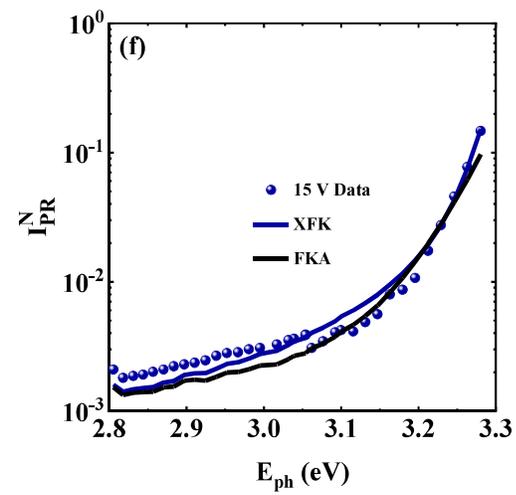

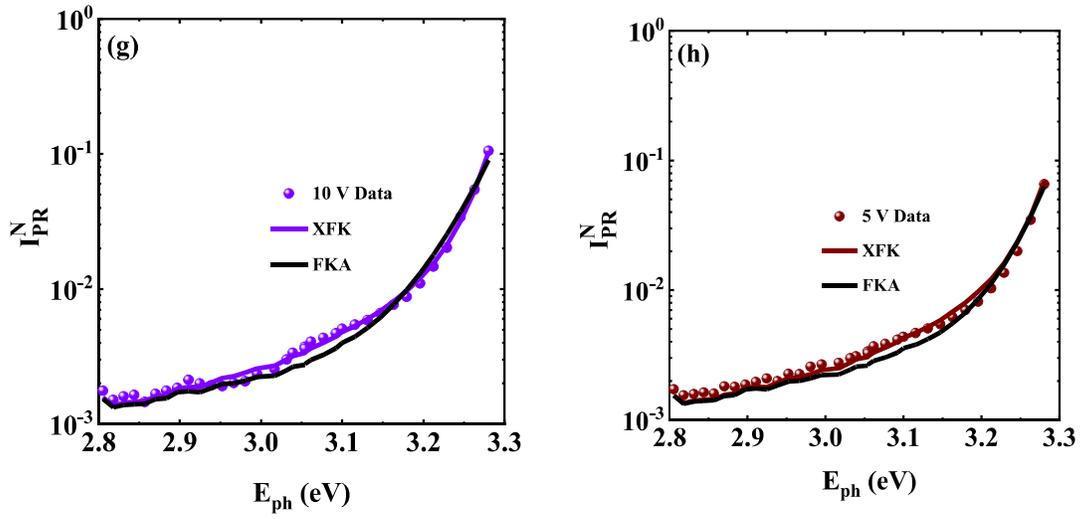

*Figure S4. (a-h) $I_{PR}^N$ spectral data (points) fit to the exciton FK (XFK) theory of Merkulov (colored lines) or the non-excitonic FKA theory (black lines) at various values of applied bias, V.*

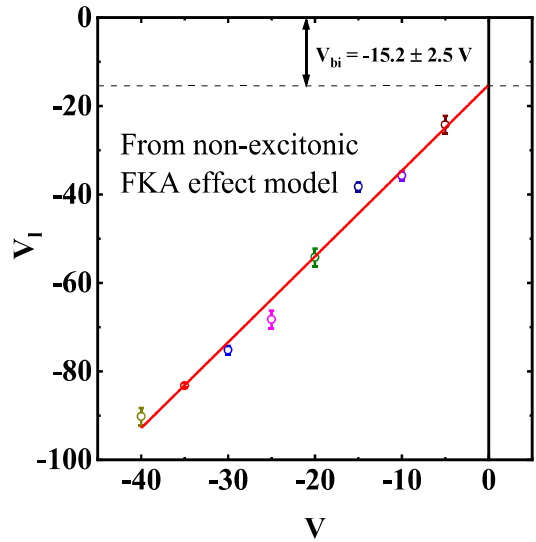

*Figure S5. Best fit values of $V_l$ determined from fits of $I_{PR}^N$ spectra to the non-excitonic FKA theory.*

## Measurement and spectral fitting reproducibility

Photocurrent responsivity spectra acquired at a second position (P2) on a different sized device (diameter ~50μm) but from the same MBE grown epilayer are shown in Figure S6(a) below. As shown in Figure S6(b), the same fitting procedure as described in the main text is performed for spectra at $V < 0$. Fit results obtained $V_l$ versus $V$ using the XFK model are shown in Figure S7. $V_{bi}$ matches the expected value within $\pm 0.7\ V$.

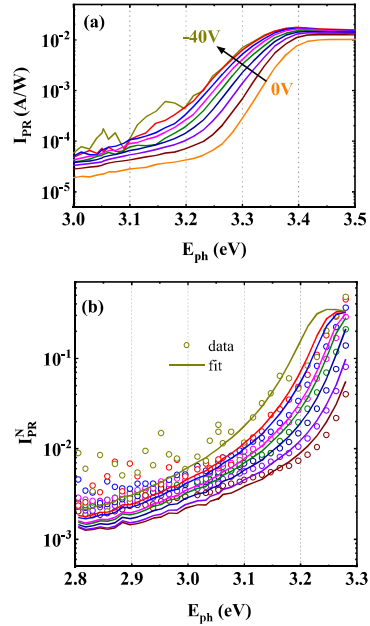

Figure S6. (a) $I_{PR}$ spectral data acquired on a second device at various values of applied bias, $V$, (b) Normalized Photocurrent Responsivities $I_{PR}^N$ best fit to XFK model

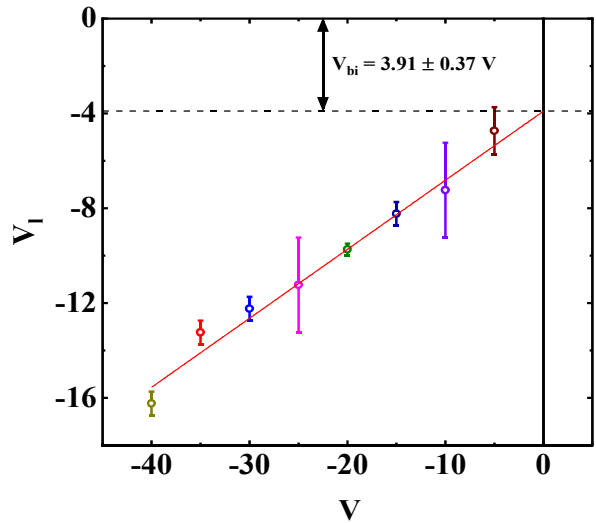

Figure S7. $V_l$ determined from best fits of the data from Fig. S6 to the XFK model.

## C-V Measurements

The $1/C^2$-V plot is shown below as well as the C-V profiling plot of $N_d$-$N_a$.

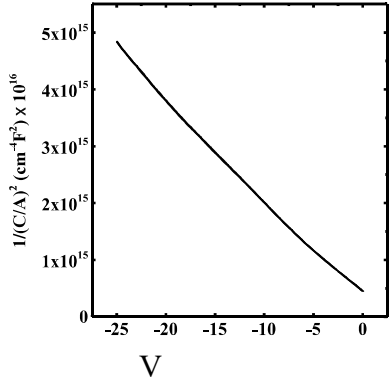
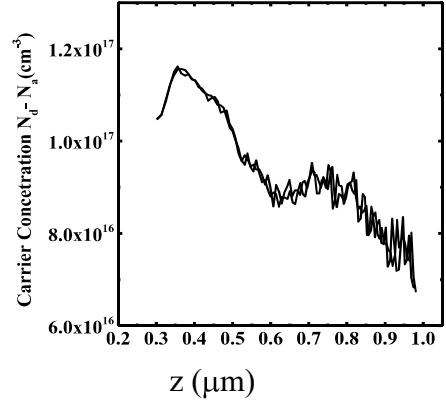

*Figure S8. C-V data.*     *Figure S9. C-V profiling data.*

SIMS is used to calibrate the Si and Mg (donor and acceptor) doping levels for MBE growth, which are $1\times10^{17}$ cm$^{-3}$ and $1\times10^{18}$ cm$^{-3}$, respectively. The C-V measured $N_d$-$N_a$ = $8\times10^{16}$ cm$^{-3}$, thus compensating $N_a$ are $2\times10^{16}$ cm$^{-3}$. The compensation is from background C acceptors in the MBE system.

---

## References

[1] I. Merkulov, Sov. J. Exp. Theor. Phys. **66**, 2314 (1974).

[2] O. Ambacher, W. Rieger, P. Ansmann, H. Angerer, T.D. Moustakas, and M. Stutzmann, Solid State Commun. **97**, 365 (1996).

[3] J.F. Muth, J.H. Lee, I.K. Shmagin, R.M. Kolbas, H.C. Casey, B.P. Keller, U.K. Mishra, and S.P. DenBaars, Appl. Phys. Lett. **71**, 2572 (1997).

[4] D.E. Aspnes, Phys. Rev. **147**, 554 (1966).

[5] T. Maeda, M. Okada, M. Ueno, Y. Yamamoto, M. Horita, and J. Suda, Appl. Phys. Express **9**, 91002 (2016).

[6] T. Maeda, X. Chi, M. Horita, J. Suda, and T. Kimoto, Appl. Phys. Express **11**, (2018).

[7] F. Binet, J. Duboz, E. Rosencher, F. Scholz, and V. Härle, Phys. Rev. B - Condens. Matter Mater.

Phys. **54**, 8116 (1996).

[8] J.Y. Duboz, F. Binet, E. Rosencher, F. Scholz, and V. Härle, Mater. Sci. Eng. B **43**, 269 (1997).

[9] M.A. Jacobson, O. V Konstantinov, D.K. Nelson, S.O. Romanovskii, and Z. Hatzopoulos, J. Cryst. Growth **230**, 459 (2001).